\documentclass[preprint,12pt,3p]{elsarticle}

\usepackage{amssymb}

\usepackage{algorithm}
\usepackage{algorithmic}

\usepackage{amsmath}

\usepackage{array,multirow,graphicx} 
\usepackage{colortbl}
\usepackage[table]{xcolor} 

\usepackage{setspace}

\spacing{1.5}

\journal{Biomedical Informatics}

\begin{document}

\begin{frontmatter}

\title{Relief-Based Feature Selection: Introduction and Review}

\author[label1]{Ryan J. Urbanowicz\corref{cor1}}
\address[label1]{Institute for Biomedical Informatics, University of Pennsylvania, Philadelphia, PA 19104, USA}

\cortext[cor1]{Corresponding Author}

\ead{ryanurb@upenn.edu}

\author[label2]{Melissa Meeker}
\address[label2]{Ursinus College, Collegeville, PA, 19426, USA }

\ead{memeeker@ursinus.edu}

\author[label1]{William LaCava}
\ead{lacava@upenn.edu}

\author[label1]{Randal S. Olson}
\ead{olsonran@upenn.edu}

\author[label1]{Jason H. Moore}
\ead{jhmoore@upenn.edu}

\begin{abstract}
Feature selection plays a critical role in biomedical data mining, driven by increasing feature dimensionality in target problems and growing interest in advanced but computationally expensive methodologies able to model complex associations. Specifically, there is a need for feature selection methods that are computationally efficient, yet sensitive to complex patterns of association, e.g. interactions, so that informative features are not mistakenly eliminated prior to downstream modeling. This paper focuses on Relief-based algorithms (RBAs), a unique family of filter-style feature selection algorithms that have gained appeal by striking an effective balance between these objectives while flexibly adapting to various data characteristics, e.g. classification vs. regression. First, this work broadly examines types of feature selection and defines RBAs within that context. Next, we introduce the original Relief algorithm and associated concepts, emphasizing the intuition behind how it works, how feature weights generated by the algorithm can be interpreted, and why it is sensitive to feature interactions without evaluating combinations of features. Lastly, we include an expansive review of RBA methodological research beyond Relief and its popular descendant, ReliefF. In particular, we characterize branches of RBA research, and provide comparative summaries of RBA algorithms including contributions, strategies, functionality, time complexity, adaptation to key data characteristics, and software availability. 
\end{abstract}


\end{frontmatter}


\section{Background} \label{intro}
The fundamental challenge of almost any data mining or modeling task is to identify and characterize relationships between one or more features in the data (also known as predictors or attributes) and some endpoint (also known as the dependent variable, class, outcome, phenotype, or concept). In most datasets, only a subset of available features are \emph{relevant features}, i.e. informative in determining the endpoint value. The remaining \emph{irrelevant features}, which are rarely distinguishable \emph{a priori} in real world problems, are not informative yet contribute to the overall dimensionality of the problem space. This increases the difficulty and computational burden placed on modeling methods. \emph{Feature selection} could generically be defined as the process of identifying relevant features and discarding irrelevant ones.  

\begin{figure}[t]
	\centering
	\includegraphics[width=\textwidth]{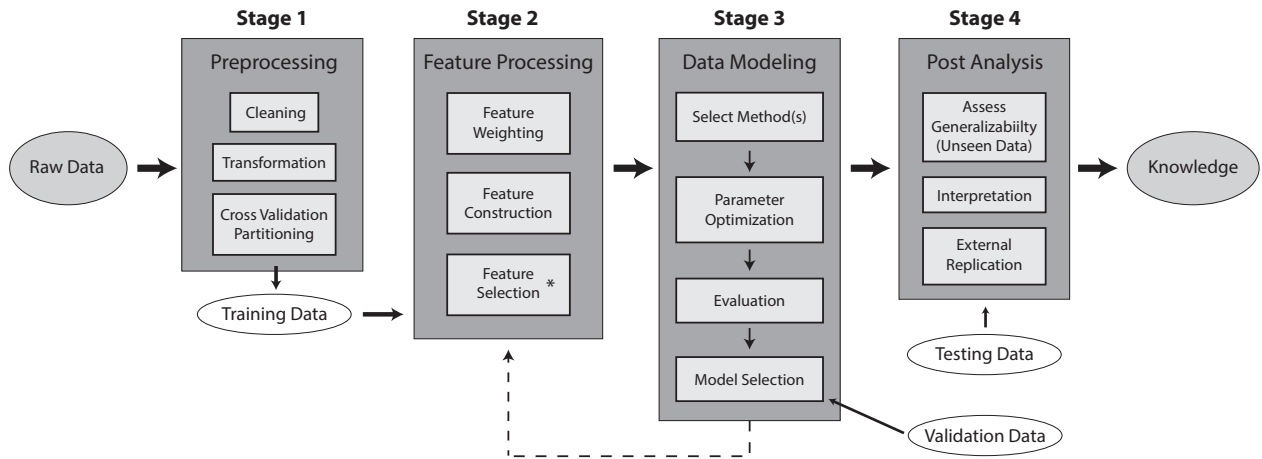} 
	\caption{Typical stages of a data mining analysis pipeline. Feature selection is starred as it is the focus of this review. The dotted line indicates how model performance can be fed back into feature processing, iteratively removing irrelevant features or seeking to construct relevant ones.}
	\label{fig:FS}
\end{figure}

Figure \ref{fig:FS} illustrates the typical stages of a data mining analysis pipeline. Specifically, raw data is preprocessed in preparation for analysis. This typically includes some type of cross validation where the data is split into training, validation, and testing subsets to avoid overfitting and assess the generalizability of the final model. Next, different feature processing approaches can be employed to remove irrelevant features or construct better relevant ones. Modeling then takes place on this preprocessed data. Model performance could then feed back into another round of feature processing (dotted line). This is the case for \emph{wrapper} feature selection methods, reviewed below. The final model is ultimately assessed and interpreted in a post analysis stage that ideally leads to the discovery of useful knowledge. Feature selection is an important part of a successful data mining pipeline, particularly in problems with very large feature spaces. Poorly performed feature selection can have significant downstream consequences on data mining, particularly when relevant features have been mistaken as irrelevant and removed from consideration.

\subsection{Types of Feature Selection} \label{intro:FS}
A large variety of feature selection methodologies have been proposed and research continues to support the claim that there is no universal ``best'' method for all tasks \citep{bolon2013review}. In order to navigate methodological options and assist in selecting a suitable method for a given task it is useful to start by characterizing and categorizing different feature selection methods \citep{dash1997feature,ni2012review,bolon2013review,jovic2015review}. One such characterization is with regards to the feature selection objective.  

\begin{enumerate}
    \item \emph{Idealized}: find the minimally sized feature subset that is necessary and sufficient to describe the target concept \citep{kira1992practical}.
    \item \emph{Target Feature Count}: select a subset of $m$ features from a total set of $n$ features, $m$ $<$ $n$, such that the value of a criterion function is optimized over all subsets of size $m$ \citep{narendra1977branch}.
    \item \emph{Prediction Accuracy Improvement}: choose a subset of features that best increases prediction accuracy or decreases model complexity without significantly decreasing the prediction accuracy \citep{inza2000feature}.
    \item \emph{Approximate Original Class Prediction Probability Distribution}: for classification problems, select a feature subset that yields a class prediction probability distribution that is as close as possible to the class prediction probability distribution given all features. In contrast with prediction accuracy this perspective seeks to preserve additional information regarding probabilities of class predictions \citep{koller1996toward}.
    \item \emph{Rank and Define Cutoff}: first rank all features using some surrogate measure of feature `value', then define the feature subset by applying an ad-hoc cutoff. This cutoff may be determined by statistical or subjective likelihood of relevance or simply a desired number of features in the subset \citep{kira1992feature}.
\end{enumerate}

\noindent This list is an updated version of one regularly used in the literature \citep{dash1997feature,mlambosurvey,tang2014feature,gore2016feature}. Alternatively, feature selection methods can be distinguished based on their relationship with the construction of the model (i.e. induction) \citep{saeys2007review,bolon2013review,chandrashekar2014survey,tang2014feature,jovic2015review,mlambosurvey}.  

\begin{enumerate}
    \item \emph{Filter Methods}: use a `proxy measure' calculated from the general characteristics of the training data to score features or feature subsets as a processing step prior to modeling.  Filters are generally much faster and function independently of the induction algorithm, meaning that selected features can then be passed to any modeling algorithm. Filter methods can be roughly classified further by the filtering measures they employ, i.e. information, distance, dependence, consistency, similarity, and statistical measures \citep{dash1997feature,bolon2013review,jovic2015review}. Examples include information gain \citep{hunt1966experiments}, chi-square \citep{jin2006machine}, and Relief \citep{kira1992practical}. 
    \item \emph{Wrapper Methods}: employ any stand-alone modeling algorithm to train a predictive model using a candidate feature subset. The testing performance on a hold-out set is typically used to score the feature set. Alternatively in a modeling algorithm like a random forest, estimated feature importance scores can be applied to select a feature subset \citep{menze2009comparison}. In any wrapper method, a new model must be trained to test any subsequent feature subset, therefore wrapper methods are typically iterative and computationally intensive, but can identify the best performing features set for that specific modeling algorithm \citep{guyon2003introduction,bolon2013review,jovic2015review}. Each iteration of the wrapper, the feature subset is generated based on the selected search strategy, e.g. forward or backward selection \citep{kittler1978feature,langley1994selection} or a heuristic feature subset selection \citep{van1987simulated,holland1992adaptation}. Examples include wrappers for Na\"ive Bayes \citep{cortizo2006multi}, Support Vector Machines (SVM) \citep{bradley1998feature}, and most any modeling algorithm combined with a feature subset generation approach. Thus a wrapper method is defined by both the selected induction algorithm as well as the feature subset search strategy. However, due to the computational complexity of wrappers, only the simplest modeling methods can be used efficiently. 
    \item \emph{Embedded Methods}: perform feature selection as a part of the modeling algorithm's execution. These methods tend to be more computationally efficient than wrappers because they simultaneously integrate modeling with feature selection. This can be done, for instance, by optimizing a two-part objective function with (1) a goodness-of-fit term and (2) a penalty for a larger number of features. As with wrappers, the features selected by embedded methods are induction algorithm dependent \citep{guyon2003introduction,bolon2013review,jovic2015review}. Examples include Lasso \citep{tibshirani1996regression}, Elastic Net \citep{zou2005regularization}, and various decision tree based algorithms, e.g. CART \citep{breiman1984classification}, C4.5 \citep{quinlan1993c4}, and most recently, XGBoost \citep{chen2016xgboost}.
\end{enumerate}

\noindent Many hybrid methods have also been proposed that seek to combine the advantages of wrappers and filters \citep{jovic2015review}.

Lastly, feature selection approaches have also been broadly categorized as relying on either \emph{individual evaluation} or \emph{subset evaluation} \citep{yu2004efficient,bolon2013review}. Individual evaluation, (i.e. feature weighting/ranking) assesses individual features and assigns them weights/scores according to their degrees of relevance \citep{blum1997selection,yu2004efficient}. Subset evaluation instead assesses candidate feature subsets that are selected based on a given search strategy \citep{bolon2013review}. Filter, wrapper, or embedded methods can be either subset or individual evaluation methods.  

The remainder of this paper will focus on the family of \emph{Relief-based} feature selection methods referred to here as Relief-Based Algorithms (RBAs) that can be characterized as \emph{individual evaluation filter methods}. For reviews of features selection methods in general, we refer readers to \citep{langley1994selection,dash1997feature,guyon2003introduction,belanche2011review,bolon2013review,tang2014feature,chandrashekar2014survey,jovic2015review,mlambosurvey}.

\subsection{Why Focus on Relief-based Feature Selection?} \label{intro:relief}
One advantage of certain wrapper or embedded methods is that by relying on subset evaluation they have the potential to capture feature dependencies in predicting the endpoint, i.e. interactions \citep{bolon2013review}. In contrast, very few filter methods, besides for example, FOCUS \citep{almuallim1991learning} and INTERACT \citep{zhao2009searching} that are notably subset evaluation filters, claim to be able to handle feature interactions. The most reliable but naive approach for identifying feature interactions is to exhaustively search over all subsets of the given feature set, e.g. FOCUS. This quickly becomes computationally intractable in problems with larger feature spaces. The inefficiency of these approaches is that they must explicitly search through combinations of features. Alternatively, the Relief algorithm, and its derivatives are, to the best of our knowledge, the only individual evaluation filter algorithms capable of detecting feature dependencies. These algorithms do not search through feature combinations, but rather use the concept of nearest neighbors to derive feature statistics that indirectly account for interactions. Furthermore, RBAs retain the generalized advantages of filter algorithms, i.e. they are relatively fast (with an asymptotic time complexity of $\mathcal{O}(instances^2\cdot features)$), and selected features are not induction algorithm dependent. The ability to confidently utilize selected features with different induction algorithms may save further downstream computational effort when applying more than a single modeling technique. This is relevant considering the 'no-free lunch' theorem proposed by \citet{wolpert1997no} that suggests no one modeling algorithm can be optimal for all problems, and the widely acknowledged value of ensemble methods reviewed by \citet{rokach2010ensemble} that combine input from multiple statistical or machine learning induction methods to make the best informed predictions. Lastly, individual evaluation approaches, including RBAs, offer a greater flexibility of use. Specifically, individual feature weights may be applied not only to select `top' features, but can also be applied as expert knowledge to guide stochastic machine learning algorithms such as evolutionary algorithms \citep{urbanowicz2012using}. Furthermore, when selecting features, feature sets of different sizes can be selected based on whatever criteria is desired for feature inclusion from a ranked feature list. 

The following two subsections address important considerations related to our assertion that RBAs deserve particular attention. A closer look at the strengths and weaknesses of the Relief algorithm is given in Section \ref{snl}.

\subsubsection{Feature Construction}
An alternative or supplemental approach to facilitate the detection and modeling of interactions is to apply feature construction (see Figure \ref{fig:FS}), also known as constructive induction or feature extraction. Feature construction methods, e.g. principle component analysis or linear discriminant analysis \citep{martinez2001pca}, define new features as a function of two or more other features \citep{michalski1983theory}. This subset of constructed features can be added to the original feature space, or analyzed in its place (achieving dimensionality reduction). A common side effect of most any feature construction method is that the original features are no longer recognizable, leading to challenges in downstream model interpretability.

One feature construction method geared specifically towards capturing feature interactions is multifactor dimensionality reduction (MDR) \citep{ritchie2001multifactor}. Another more general example is polynomial feature construction that is able to detect multiplicative interactions \citep{sutton1991learning}. These approaches attempt to combine individual features that may be interacting and construct a single feature that can be more easily identified as relevant using any simple feature selection or induction method. There are many possible feature construction approaches to chose from and some can be quite computationally expensive. Notably, applying feature construction does not necessarily preclude the need for feature selection. Thus, assuming that a feature selection and modeling approach has been chosen that is sensitive to a target interaction dimensionality (e.g. 2-way or 3-way), it may be most efficient to skip feature construction, particularly if downstream model interpretation is critical. While feature construction certainly has its own utility, further discussion is outside the scope of this review. 

\subsubsection{Redundancy}
Relevant features can be more restrictively defined as any feature that is neither irrelevant nor redundant to the target concept \citep{koller1996toward,dash1997feature}. Feature redundancy is explored further by \citet{yu2004efficient}. Some feature selection methods seek to remove redundant features while others do not. Caution should be used when removing presumably redundant features, because unless two features are perfectly correlated (i.e. truly redundant) there may still be information to be gained from including them both \citep{guyon2003introduction}.
One repeatedly noted drawback of RBAs is that they do not remove feature redundancies, i.e. they seek to select all features relevant to the endpoint regardless of whether some features are strongly correlated with others \citep{kira1992practical,belanche2011review, florez2002reviewing}. However, except for features that are perfectly correlated, it is not always clear whether useful information is being lost when `redundant' features are removed. For example, it has been suggested that preserving redundant features can be a benefit, as it ``may point to meaningful clusters of correlated phenotypes'' \citep{todorov20166}. If removing redundancy is clearly important to success in a given problem domain, many effective methods are available that can be applied before, after, or integrated with RBA feature selection to remove feature redundancies \citep{bins2001feature,florez2002reviewing,yang2006orthogonal,sun2007iterative,challita2015new,agre2016weighted,liu2015feature,guyon2003multivariate}.  

\subsection{Paper Summary}
In the text that follows, we (1) introduce RBAs from the perspective of the original Relief algorithm noting key concepts and intuitions, (2) examine the contributions of the landmark ReliefF algorithm, (3) differentiate thematically distinct branches of RBA research, (4) review methodological expansions and advancements introduced by derivative members of the RBA family in the wake of Relief and ReliefF, (5) consider RBA evaluations, and (6) summarize software availability.  This review was prepared to complement a comprehensive research comparison of `core' RBAs presented by \citet{urbanowicz2017rebate}. 

\section{Introduction to Relief} \label{back}
In this section we provide algorithmic and conceptual descriptions of the original Relief algorithm relevant to understanding all members of the RBA family. 

\subsection{Relief} \label{sec:relief}
\citet{kira1992practical,kira1992feature} formulated the original Relief algorithm inspired by instance-based learning \citep{aha1991instance,callan1991cabot}. As an individual evaluation filtering feature selection method, Relief calculates a proxy statistic for each feature that can be used to estimate feature `quality' or `relevance' to the target concept (i.e. predicting endpoint value). These feature statistics are referred to as feature weights ($W[A] =$ weight of feature `$A$'), or more casually as feature `scores' that can range from $-1$ (worst) to $+1$ (best). Notably, the original Relief algorithm was limited to binary classification problems, and had no mechanism to handle missing data. Strategies to extend Relief to multi-class or continuous endpoint problems are not detailed here, but are described in the respective works cited in the review section of this paper.

\begin{algorithm}[h]
\caption{Pseudo-code for the original Relief algorithm}
\label{alg:Relief}
\begin{algorithmic}
\REQUIRE for each training instance a vector of feature values and the class value
\STATE $\textit{n} \gets$ number of training instances
\STATE $\textit{a} \gets$ number of features (i.e. attributes)
\STATE \textbf{Parameter:} $\textit{m} \gets$ number of random training instances out of $n$ used to update $W$ \\
\STATE
\STATE initialize all feature weights $W[A]:=0.0$
\FOR{i:=1 \TO $m$} 
    \STATE randomly select a `target' instance $R_{i}$
    \STATE find a nearest hit `$H$' and nearest miss `$M$' (instances)
    \FOR{A:= 1 \TO $a$} 
        \STATE $W[A]:= W[A] - $\emph{diff}$(A,R_{i},H)/m+$\emph{diff}$(A,R_{i},M)/m$
    \ENDFOR
\ENDFOR
\RETURN the vector $W$ of feature scores that estimate the quality of features
\end{algorithmic}
\end{algorithm}

\begin{figure}[t]
	\centering
    \includegraphics[width=0.65\textwidth]{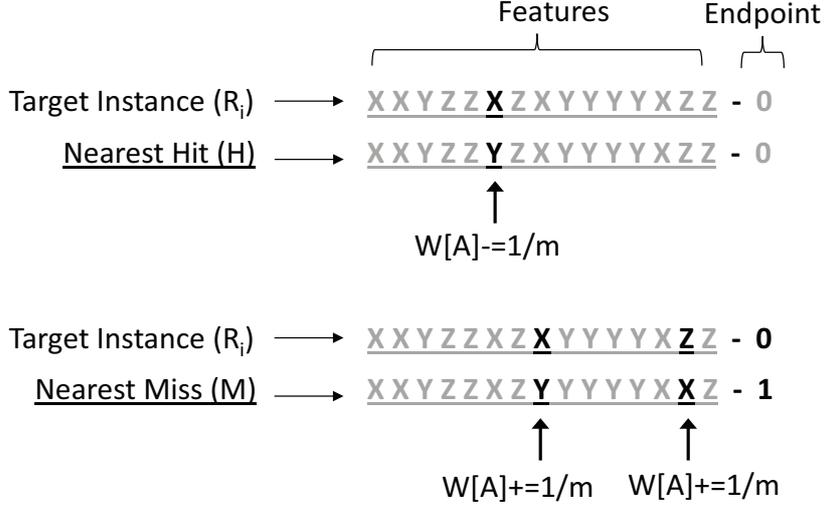}
	\caption{Relief updating $W[A]$ for a given target instance when it is compared to its nearest miss and hit. In this example, features are discrete with possible values of X, Y, or Z, and endpoint is binary with a value of 0 or 1. Notice that when the value of a feature is different, the corresponding feature weight increases by $1/m$ for the nearest miss, and reduces by $1/m$ for the nearest hit.}
	\label{fig:RS}
\end{figure}

As summarized by the pseudo-code in Algorithm \ref{alg:Relief}, the Relief algorithm cycles through $m$ random training instances ($R_{i}$), selected  without replacement, where $m$ is a user-defined parameter. Each cycle, $R_{i}$ is the `target' instance and the feature score vector \emph{W} is updated based on feature value differences observed between the target and neighboring instances. Therefore each cycle, the distance between the `target' instance and all other instances is calculated. Relief identifies two nearest neighbor instances of the target; one with the same class, called the \emph{nearest hit} ($H$) and the other with the opposite class, called the \emph{nearest miss} ($M$). The last step of the cycle updates the weight of a feature $A$ in $W$ if the feature value differs between the target instance $R_{i}$ and either the nearest hit $H$ or the nearest miss $M$ (see Figure \ref{fig:RS}). Features that have a different value between $R_{i}$ and $M$ support the hypothesis that they are informative of outcome, so the quality estimation \emph{W}[\emph{A}] is increased. Conversely, features with differences between $R_{i}$ and $H$ provide evidence to the contrary, so the quality estimation \emph{W}[\emph{A}] is decreased. The \emph{diff} function in Algorithm \ref{alg:Relief} calculates the difference in value of feature $A$ between two instances $I_{1}$ and $I_{2}$, where $I_{1}=R_{i}$ and $I_{2}$ is either $H$ or $M$, when performing weight updates \citep{robnik2001comprehensible}. For discrete (e.g. categorical or nominal) features, \emph{diff} is defined as:

\begin{equation}\label{eq:disc}
\text{\emph{diff}}(A,I_{1},I_{2}) =
\begin{cases}
0 & \text{if } value(A,I_{1})=value(A,I_{2})\\
1 & \text{if } otherwise\\
\end{cases}
\end{equation} 

\noindent and for continuous (e.g. ordinal or numerical) features, \emph{diff} is defined as:

\begin{equation}\label{eq:cont}
\text{\emph{diff}}(A,I_{1},I_{2}) = \frac{|value(A,I_{1})-value(A,I_{2})|}{max(A)-min(A)}
\end{equation} 

\noindent The maximum and minimum values of A are determined over the entire set of instances. This normalization ensures that weight updates fall between 0 and 1 for both discrete and continuous features. Additionally, in updating \emph{W}[\emph{A}], dividing the output of \emph{diff} by $m$ guarantees that all final weights will be normalized within the interval $[-1,1]$. 

The \emph{diff} function is also used to calculate the distance between instances when finding nearest neighbors. The total distance is simply the sum of \emph{diff} distances over all attributes (i.e. Manhattan distance). Technically, the original Relief algorithm used Euclidian distance rather than Manhattan distance i.e. the \emph{diff} terms were squared during instance distance measurements and feature weighting. However, experiments by \citep{kononenko1997overcoming} indicated no significant difference between results using \emph{diff} or squared \emph{diff}, thus the simplified description of the Relief algorithm has become standard. It has also been suggested that any valid distance metric could be used by Relief \citep{todorov20166}. Thus, determining the best distance metric remains an open research question. While the above \emph{diff} function performs well when features are either uniformly discrete or continuous, it has been noted that given a dataset with a mix of discrete and continuous features, this \emph{diff} function can underestimate the quality of the continuous features \citep{kononenko2008non}. One proposed solution to this problem is a \emph{ramp function} that naively assigns a full \emph{diff} of 0 or 1 if continuous feature values are some user defined minimum or maximum value apart from one another, respectively, and a function of the distance from these boundaries otherwise \citep{hong1997use,robnik2003theoretical,kononenko2008non}. However since this approach adds two additional user-defined parameters requiring problem dependent optimization, it may be challenging to apply in practice.

\subsubsection{Strengths and Limitations} \label{snl}
Regarding strengths, Relief has been presented as being both non-myopic \citep{kononenko2008non}, i.e. it estimates the quality of a given feature in the context of other features, and non-parametric \citep{todorov20166}, i.e. it makes no assumptions regarding the population distribution or sample size. The efficiency of the algorithm has been attributed to the fact that it doesn't explicitly explore feature subsets and because it does not bother trying to identify an optimal minimum feature subset size \citep{kira1992practical}. Instead, Relief was originally ``intended as a \emph{screener} to identify a subset of features that may not be the smallest and may still include some irrelevant and redundant features, but that is small enough to use with more refined approaches in a detailed anlaysis'' \citep{todorov20166}. Consider that an exhaustive search for interactions between all feature pairs alone would have a time complexity of $\mathcal{O}(2^{a})$, while Relief boasts a time complexity of $\mathcal{O}(a \cdot m \cdot n)$, or $\mathcal{O}(a \cdot n)$ whenever $m < n$. Furthermore, it has been suggested that Relief could be viewed as an \emph{anytime algorithm}, i.e. one that can be stopped and yield results at any time, but it is presumed that with more time or data it will improve the results \citep{robnik2003theoretical}.

Regarding limitations, the original Relief analysis suggests that the algorithm can be fooled by insufficient training cycles (i.e. not a large enough $m$). The original paper also suggests that Relief is fairly noise-tolerant and unaffected by feature interactions. However, later work identified that Relief was susceptible to noise interfering with the selection of nearest neighbors \citep{kononenko1994estimating}. Further, research into RBAs has, until recently, been limited to considering 2-way feature interactions only. Therefore, it was unclear if RBAs could detect feature interactions with a dimensionality beyond 2 features. Research paired with this review suggests that only specific RBAs have the ability to detect higher order interactions (e.g. 3-way, 4-way, and 5-way), thus RBAs are only universally reliable in detecting 2-way interactions \citep{urbanowicz2017rebate}. Relief has also been noted to have a reduced power to identify relevant non-monotonic features (e.g. features with a Gaussian distribution) \citep{bins2002evaluating}.  
Most importantly, it has been repeatedly demonstrated empirically and theoretically that core RBA performance deteriorates as the number of irrelevant features becomes `large'. \citep{robnik2003theoretical,moore2007tuning,eppstein2008very,todorov20166}. This deterioration of performance in identifying interacting features is primarily due to the fact that Relief's computation of neighbors and weights becomes increasingly random as the number of features increases. This is an example of the curse of dimensionality. The iterative RBAs reviewed in Section \ref{sec:iterative} have been demonstrated to improve RBA performance in these types of large feature spaces. Differently, deteriorating performance in detecting main effects in very large feature spaces is primarily due to feature scores being based on feature value differences between a subset of neighboring instances, rather than differences from all instances. Thus the main effect signal in RBA scores is not expected to be as pronounced. Given a very large feature space, this less pronounced main effect feature score is less likely to stand out. In general, it is likely that myopic feature selection algorithms that compute scores by comparing the feature values of all training instances will have the most power to detect simple main effects \citep{mckinney2013reliefseq}. In addition to an iterative RBA approach, another way to address lost main effect performance could involve running both an RBA and a myopic feature selection algorithm, selecting the top non-redundant set of features combined from both algorithms. Unfortunately, as of yet, there are no clear guidelines regarding the size of the feature space where (1) myopic methods would be expected to outperform RBAs in detecting main effects, or (2) interactions effects can't be distinguished from random background noise. Simulations studies such as those reviewed in Section \ref{evals} offer some insight, however in real-world applications many factors are expected impact success (e.g. number of training instances, type of signal, signal strength, feature distributions, feature type, etc.).

Lastly, it's notable that Relief scores do not reflect the nature of an association. For example Relief does not tell you which attributes might be interaction partners, or whether a score is high due to a linear effect or an interaction. This is left to downstream modeling. Furthermore, there is no established way to assess how many of the high scoring selected features may be false discoveries. It is possible this issue could be addressed through permutation testing as suggested by \citet{mckinney2013reliefseq}.

\subsection{Feature Subset Selection}
The original description of Relief specified an automated strategy for feature subset selection \citep{kira1992practical}. Specifically, a \emph{relevance threshold} ($\tau$) was defined such that any feature with a relevance weight $W$[$A$] $\geq \tau$ would be selected. Kira and Rendell demonstrated that ``statistically, the relevance level of a relevant feature is expected to be larger than zero and that of an irrelevant one is expected to be zero (or negative)''. Therefore generally the threshold should be selected such that $0 < \tau < 1$. More precisely they proposed the bounds i.e. $0 < \tau \leq \frac{1}{\sqrt{\alpha m}}$, based on Chebyshev's inequality, where $\alpha$ is the probability of accepting an irrelevant feature as relevant (i.e. making a Type I error). If $\tau$ is set too high, there is an increased chance that one or more relevant features will fail to be selected. Alternatively if $\tau$ is set too low, it is expected that an increased number of irrelevant features will be selected. Like any significance threshold the choice of $\tau$ is somewhat arbitrary. Not all features with a weight above the selected threshold will necessarily be relevant because it is expected that some irrelevant features will have a positive weight by chance. 

In practice, rather than choosing a value of $\tau$, it is often more practical to choose some number of features to be selected \emph{a priori} based on the functional, computational, or run time limitations of the downstream modeling algorithms that will be applied. Ultimately the goal is to provide the best chance that all relevant features are included in the selected set for modeling, but at the same time, remove as many of the irrelevant features as possible to facilitate modeling, reduce over-fitting, and make the task of induction tractable.

\subsection{Intuition, Interpretation, and Interactions} \label{sec:int}
Relief algorithms often appear simple at first glance, but understanding how to interpret feature weights and gaining the intuition as to how feature dependencies (i.e. interactions) can be gleaned without explicitly considering feature subsets is not always apparent. The key idea behind Relief is to estimate feature relevance according to how well feature values distinguish concept (endpoint) values among instances that are similar (near) to each other. Two complementary interpretations of Relief feature weights have been derived and presented: (1) a probabilistic interpretation \citep{kononenko1994estimating,kononenko1996relieff,kononenko1997overcoming,robnik2001comprehensible,robnik2003theoretical} and (2) an interpretation as the portion of the explained concept changes \citep{robnik2001comprehensible,robnik2003theoretical}.  Next, we summarize these interpretations and why they explain Relief's ability to detect interactions. 

\subsubsection{Probabilistic Interpretation}
The first interpretation is that the Relief weight estimate $W$[$A]$ of feature $A$ is an approximation of the following difference of probabilities:

\begin{equation}\label{eq:weight}
\begin{split}
W[A] &= P(\text{different value of $A$ } | \text{ nearest instance from different class}) \\
&- P(\text{different value of $A$ } | \text{ nearest instance from same class})
\end{split}
\end{equation} 

\noindent Consider that as the number of nearest neighbors used in scoring increases from 1 and approaches $n$ this effectively eliminates the condition that instances used in scoring be `near'. Notably if we were to eliminate the `near' requirement from  \ref{eq:weight}, the formula becomes:

\begin{equation}\label{eq:weight_alt}
\begin{split}
W[A] &= P(\text{different value of $A$ } | \text{ different class}) \\
&- P(\text{different value of $A$ } | \text{ same class})
\end{split}
\end{equation} 

\noindent As derived by \citet{robnik2003theoretical} it can be shown that, without the near condition, Relief weights would be strongly correlated with impurity functions such as the Gini-index gain.  Impurity functions including information gain \citep{hunt1966experiments}, gain ratio \citep{quinlan1993c4}, gini-index \citep{breiman1984classification}, distance measure \citep{de1991distance}, j-measure \citep{smyth1992information}, and MDL \citep{kononenko1995biases} have often been used as myopic filter feature selection algorithms that assume features to be conditionally independent given the class.  

Thus, it is the `\emph{nearest instance}' condition in Equation \ref{eq:weight} and the resulting fact that Relief weights are averaged over local estimates in smaller parts of the instance subspace (rather than globally over all instances) that enables Relief algorithms to take into account the context of other features and detect interactions \citep{kononenko1997overcoming,kononenko2008non}. It has been demonstrated by \citet{robnik2003theoretical}, that as the number of neighbors used in scoring approaches $n$ the ability of Relief to detect feature dependencies disappears, since scoring is no longer limited to `near' instances. 

\subsubsection{Concept Change Interpretation}
The second interpretation of Relief weights has been argued as being the more comprehensible/communicable than the probabilistic one \citep{robnik2001comprehensible}. The authors demonstrate that Relief relevance weights $W$[$A]$ can be interpreted as the ratio between the number of explained changes in the concept and the number of examined instances. If a particular change can be explained in multiple ways, all ways share credit for it in the quality estimate. Also if several features are involved in one way of the explanation, all of them get the credit in their quality estimate \citep{robnik2003theoretical}.  To illustrate this idea, Table \ref{tab:Bool} presents a simple Boolean problem where $Class$ is determined by the expression ($A_{1} \wedge A_{2}$) $\vee$ ($A_{1} \wedge A_{3}$), such that all three features ($A_{1}$, $A_{2}$, and $A_{3}$) are relevant.

\begin{table*}[h]
 \centering
\caption{Tabular dataset description of Boolean problem $Class$ = ($A_{1} \wedge A_{2}$) $\vee$ ($A_{1} \wedge A_{3}$) including the responsibility of each feature for yielding an expected class change. Adapted from \citet{robnik2003theoretical}.}
\label{tab:Bool}       
\begin{tabular}{|c||ccc|c||c||ccc|}
\hline\
&\multicolumn{3}{c|}{Feature Values} &&& \multicolumn{3}{c|}{Score Change} \\
Instances & $A_{1}$ & $A_{2}$ & $A_{3}$ & $Class$ & Responsible Features & $A_{1}$ & $A_{2}$ & $A_{3}$ \\ \hline
$R_{1}$ & 1 & 1 & 1 & 1 & $A_{1}$&$1/8$&0&0 \\
$R_{2}$ & 1 & 1 & 0 & 1 & $A_{1}$ or $A_{2}$&$0.5/8$&$0.5/8$&0 \\
$R_{3}$ & 1 & 0 & 1 & 1 & $A_{1}$ or $A_{3}$&$0.5/8$&0&$0.5/8$ \\
$R_{4}$ & 1 & 0 & 0 & 0 & $A_{2}$ or $A_{3}$&0&$0.5/8$&$0.5/8$ \\
$R_{5}$ & 0 & 1 & 1 & 0 & $A_{1}$&$1/8$&0&0 \\
$R_{6}$ & 0 & 1 & 0 & 0 & $A_{1}$&$1/8$&0&0 \\
$R_{7}$ & 0 & 0 & 1 & 0 & $A_{1}$&$1/8$&0&0 \\
$R_{8}$ & 0 & 0 & 0 & 0 & ($A_{1}$ and $A_{2}$) or &$1/8$&$0.5/8$&$0.5/8$ \\
&&&&&($A_{1}$ and $A_{3}$)&&&\\ \hline \cline{6-9}
\multicolumn{5}{c|}{} &\multicolumn{1}{c|}{Total}& 0.75 & 0.1875 & 0.1875 \\
\cline{6-9}
\end{tabular}
\end{table*}

In the first instance of Table \ref{tab:Bool} it can be said that $A_{1}$ is responsible for class assignment because changing its value would be the only feature value change necessary to make $Class=0$. In the second instance, changing either $A_{1}$ or $A_{2}$ would make $Class=0$, thus they share the responsibility. Similar responsibility assignments can be made for instances 3 to 7, while in instance 8, changing only one feature value isn't enough for $Class$ to change, however there are two pairs of feature value changes that can.  As detailed by \citet{robnik2003theoretical}, adding up the responsibility for each feature results in a score estimate of 0.75 for $A_{1}$, and an estimate of 0.1875 for both $A_{2}$ and $A_{3}$.  This result makes sense given that $A_{1}$ clearly has a stronger linear association with $Class$ (i.e. a main effect), but both $A_{2}$ and $A_{3}$ contribute to a lesser extent, interacting with $A_{1}$ in a subset of instances. This conceptual example was validated empirically by \citet{robnik2003theoretical}, finding in a dataset with these three relevant features along with five random binary features, that the Relief relevance estimate for $A_{1}$ converges near 0.75, and estimates for $A_{2}$ and $A_{3}$ converge near 0.1875 with an increasing number of training instances.

\subsubsection{Breaking Down Interaction Detection}
To further clarify how Relief detects interactions, Table \ref{tab:Epi} offers a simple example dataset that we will use to walk through Relief scoring. In this example, $A_{1}$ and $A_{2}$ are relevant features. When they have different values, the $Class=1$ otherwise the $Class=0$. This is an example of a `pure' interaction, where no individual feature has an association with endpoint. $A_{3}$ is an irrelevant feature.  

\begin{table*}[h]
 \centering
\caption{Example dataset with interaction between $A_{1}$ and $A_{2}$. $A_{3}$ is irrelevant. Adapted from \citep{kononenko1997overcoming}.}
\label{tab:Epi}       
\begin{tabular}{|c||ccc|c|}
\hline
Instances & $A_{1}$ & $A_{2}$ & $A_{3}$ & $Class$ \\ \hline
$R_{1}$ & 1 & 0 & 1 & 1   \\
$R_{2}$ & 1 & 0 & 0 & 1   \\
$R_{3}$ & 0 & 1 & 1 & 1   \\
$R_{4}$ & 0 & 1 & 0 & 1   \\
$R_{5}$ & 0 & 0 & 1 & 0   \\
$R_{6}$ & 0 & 0 & 0 & 0   \\
$R_{7}$ & 1 & 1 & 1 & 0   \\
$R_{8}$ & 1 & 1 & 0 & 0   \\

\hline
\end{tabular}
\end{table*}

Table \ref{tab:EpiB} breaks down how scoring would proceed over 8 cycles with each instance getting to be the respective target. For each target, we see what instance is the nearest hit and miss, as well as what feature has a different value between the instances (given in parentheses), and thus is relevant to scoring. If there is a tie for nearest neighbor, both instances are listed with their respective different valued feature. For example when $R_{1}$ is the target, its nearest hit is $R_{2}$. The only feature with a different value between these two instances is $A_{3}$. The nearest miss for $R_{1}$ is a tie between $R_{5}$ and $R_{7}$ that have feature value differences at $A_{1}$ and $A_{2}$, respectively.

\begin{table*}[h]
 \centering
\caption{Breakdown of Relief nearest neighbors (i.e. hits and misses) and corresponding feature value differences given in parentheses when a given instance from Table \ref{tab:Epi} is the target.}
\label{tab:EpiB}       
\begin{tabular}{|c||c|c|}
\cline{2-3}
\multicolumn{1}{c|}{} & \multicolumn{2}{c|}{Nearest}  \\ \hline
\multicolumn{1}{|c||}{Target} & \multicolumn{1}{c|}{Hit} & \multicolumn{1}{c|}{Miss}  \\ \hline
$R_{1}$& $R_{2}(A_{3})$ & $R_{5}(A_{1})$,$R_{7}(A_{2})$   \\
$R_{2}$& $R_{1}(A_{3})$ & $R_{6}(A_{1})$,$R_{8}(A_{2})$   \\
$R_{3}$ & $R_{4}(A_{3})$ &$R_{5}(A_{2})$,$R_{7}(A_{1})$   \\
$R_{4}$& $R_{3}(A_{3})$ & $R_{6}(A_{2})$,$R_{8}(A_{1})$   \\
$R_{5}$& $R_{6}(A_{3})$ & $R_{1}(A_{1})$,$R_{3}(A_{2})$   \\
$R_{6}$ & $R_{5}(A_{3})$ &$R_{2}(A_{1})$,$R_{4}(A_{2})$   \\
$R_{7}$& $R_{8}(A_{3})$ & $R_{1}(A_{2})$,$R_{3}(A_{1})$   \\
$R_{8}$& $R_{7}(A_{3})$ & $R_{2}(A_{2})$,$R_{4}(A_{1})$   \\

\hline
\end{tabular}
\end{table*}

Table \ref{tab:Episum} summarizes the resulting number of nearest hit and miss score contributions from the Table \ref{tab:EpiB}.  When there is a tie between instances for nearest neighbor, we give each feature difference half credit since only one can contribute at a time. We can see from Table \ref{tab:EpiB} that among nearest hits we observe no feature value differences for $A_{1}$ or $A_{2}$, but a total of 8 of them for $A_{3}$ across all 8 cycles. Among nearest misses, we observe 8 feature value differences for both $A_{1}$ and $A_{2}$. However since it would be only one or the other each scoring iteration they each receive a total of 4. Lastly, the Relief scoring scheme applies negative scoring to nearest hits, and positive scoring to nearest misses. As seen in this simple example, Relief easily differentiate between the relevant interacting features $A_{1}$ and $A_{2}$ (each with a final score of 4) and the irrelevant feature $A_{3}$ (with a final score of $-8$). 

\begin{table*}[h]
 \centering
\caption{Summary of score contributions in 2-way epistasis problem yielding Relief scores.}
\label{tab:Episum}       
\begin{tabular}{|c|ccc|c|}
\cline{1-5}
 & $A_{1}$ & $A_{2}$ & $A_{3}$ & Relief Scoring  \\ \hline
Nearest Hit & 0 & 0 & 8 & - 1  \\
Nearest Miss & 4 & 4 & 0 & +1  \\ \hline 
\multicolumn{1}{|c|}{Relief Score Total} & \multicolumn{1}{c|}{\cellcolor{green!15}4} & \multicolumn{1}{c|}{\cellcolor{green!15}4} & \multicolumn{1}{c|}{\cellcolor{red!30}-8} & \multicolumn{1}{c}{} \\

\cline{1-4}
\end{tabular}
\end{table*}

\section{A Review of Relief-based Algorithms}
In this section, we offer a comprehensive review of the RBAs inspired by Relief highlighting major research themes, advancements, concepts, and adaptations.

\subsection{ReliefF: The Best Known Variant}
 The original Relief algorithm \citep{kira1992practical} is rarely applied in practice anymore and has been supplanted by \emph{ReliefF} \citep{kononenko1994estimating} as the best known and most utilized RBA to date. Notably, the 'F' in ReliefF refers to the sixth algorithm variation (from A to F) proposed by \citet{kononenko1994estimating}. The ReliefF algorithm has been detailed in a number of other publications \citep{kononenko1996relieff,kononenko1997overcoming,robnik2003theoretical}. Here we highlight four key ways that ReliefF differs from Relief. First, ReliefF relies on a `number of neighbors' user parameter $k$ that specifies the use of $k$ nearest hits and $k$ nearest misses in the scoring update for each target instance (rather than a single hit and miss). This change increased weight estimate reliability, particularly in noisy problems. A $k$ of 10 was suggested based on preliminary empirical testing and has been widely adopted as the default setting. This algorithm variation was originally proposed under the name \emph{ReliefA}. 
 
 Second, three different strategies were proposed to handle incomplete data (i.e. missing data values).  These strategies were proposed under the names Relief(B-D). When encountering a missing value, the `best' approach (\emph{ReliefD}), sets the \emph{diff} function equal to the class-conditional probability that two instances have different values for the given feature. This is implicitly an interpolation approach. 
 
 Third, two different strategies were proposed to handle multi-class endpoints. These strategies were proposed under the names ReliefE and ReliefF. \emph{ReliefF}, which inherited the changes proposed in ReliefA and ReliefD, was selected as the `best' approach. During scoring in multi-class problems, ReliefF finds $k$ nearest misses from \emph{each} `other' class, and averages the weight update based on the prior probability of each class. Conceptually, this encourages the algorithm to estimate the ability of features to separate all pairs of classes regardless of which two classes are closest to one another. 
 Lastly, since it is expected that as the parameter $m$ approaches the total number of instances $n$, the quality of the weight estimates becomes more reliable, \citet{kononenko1994estimating} proposed the simplifying assumption that $m=n$. In other words, every instance in the dataset gets to be the target instance one time (i.e instances are selected without replacement). We adopt this assumption in deriving the time complexity of RBAs below. This is why the asymptotic time complexity of `core' Relief algorithms are given as $\mathcal{O}(n^2\cdot a)$, rather than $\mathcal{O}(m\cdot n\cdot a)$.

\begin{table*}[t!]

 \centering
\caption{Summary of key Relief-based algorithms.}
\label{tab:RBM}       
{\tiny 

\begin{tabular}{l|l|c|c|c|c|c}

\hline
&& \multirow{8}{*}{\rotatebox[origin=l]{90}{\textbf{Focus}}} & \multirow{8}{*}{\rotatebox{90}{\textbf{Continuous F.}}} & \multirow{8}{*}{\rotatebox{90}{\textbf{Multi-class}}} & \multirow{8}{*}{\rotatebox{90}{\textbf{Regression}}} & \multirow{8}{*}{\rotatebox{90}{\textbf{Missing Data}}} \\

\textbf{Algorithm, Reference(s), \& } & \textbf{Time Complexity} &&&&& \\
\textbf{ Description/Contribution } &&&&&& \\ 
(\textit{Closest Parent Algorithm}) & \cellcolor{yellow!50}Asymptotic ($\mathcal{O}$)&&&&& \\ 
& \cellcolor{blue!20}Complete (Approx.) &&&&&  \\ 
&&&&&& \\ 
&&&&&& \\ 
&&&&&& \\ \hline
\rowcolor{gray!30} \textbf{Relief} \citep{kira1992practical,kira1992feature} The first `filter' feature selection algorithm & \cellcolor{yellow!50}$\mathcal{O}(n^2\cdot a)$ & C & X & & &  \\ 
\rowcolor{gray!30} sensitive to feature dependencies.  & \cellcolor{blue!20}$\chi + c_{3} 2na $ & & & & &  \\

\textbf{ReliefA} \citep{kononenko1994estimating} Introduced $k$ nearest neighbor scoring to & \cellcolor{yellow!50}$\mathcal{O}(n^2\cdot a)$ & C & X & & &   \\ 
address noisy data. (\textit{Relief})& \cellcolor{blue!20}$\chi + c_{3} 2k n a $ & & & & &   \\

\rowcolor{gray!30} \textbf{Relief(B-D)} \citep{kononenko1994estimating} Strategies for handling incomplete data. & \cellcolor{yellow!50}$\mathcal{O}(n^2\cdot a)$ & D & X & & & X  \\ 
\rowcolor{gray!30} ReliefD selected as `best'. (\textit{ReliefA}). & \cellcolor{blue!20}$\chi + c_{3} 2kna $ & & & & &  \\

\textbf{Relief(E-F)} \citep{kononenko1994estimating} Strategies for multi-class endpoint. & \cellcolor{yellow!50}$\mathcal{O}(n^2\cdot a)$ & D & X & X & & X  \\ 
 ReliefF became the standard. (\textit{ReliefD})& \cellcolor{blue!20}$\chi + c_{3} 2kna $ & & & & &  \\

\rowcolor{gray!30} \textbf{RReliefF} \citep{kononenko1996relieff} Adapting to regression problems. \citep{robnik1997adaptation} & \cellcolor{yellow!50}$\mathcal{O}(n^2\cdot a)$ & D & X & & X & X  \\
\rowcolor{gray!30}Adopts exponential instance weighting by distance from & \cellcolor{blue!20}$\chi + c_{3} n a + c_{y}(1+2a)$ & C & & & &   \\
\rowcolor{gray!30}  target. (\textit{ReliefF}) & \cellcolor{blue!20}$+ c_{y1}0.5n^2 + c_{y2}kn + c_{y3}2kna$ & & & & &  \\

\textbf{Relieved-F} \citep{kohavi1997wrappers} Deterministic neighbor selection and & \cellcolor{yellow!50}$\mathcal{O}(n^2\cdot a)$ & C & X & X & & X  \\ 
 incomplete data handling. (\textit{ReliefF}) & \cellcolor{blue!20}$\chi + c_{3} 2kna$ & D & & & &  \\

\rowcolor{gray!30} \textbf{Iterative Relief} \citep{draper2003iterative} Address bias against non-monotonic & \cellcolor{yellow!50}$\mathcal{O}(\mathcal{I} \cdot n^2\cdot a)$ & C & X & & &  \\ 
\rowcolor{gray!30}  features. The first iterative approach. Neighbors uniquely & \cellcolor{blue!20}$\mathcal{I} (\chi + c_{3}0.5n^2a + c_{y1}a)$ & I & & & &   \\
\rowcolor{gray!30}determined by radius ($r$) and instances weighted by & \cellcolor{blue!20}$+c_{y2}a$  & & & & &  \\
\rowcolor{gray!30}  distance from target. (\textit{Relief})& \cellcolor{blue!20} & & & & &  \\

\textbf{I-RELIEF} \citep{sun2006iterative,sun2007iterative} All instances sigmoidally weighted by & \cellcolor{yellow!50}$\mathcal{O}(\mathcal{I} \cdot n^2\cdot a)$ & C & X & X & &  \\ 
 distance from target, i.e. no defined neighbors. Proposed  & \cellcolor{blue!20}$\mathbf{\mathcal{I}} (\chi + c_{3}n^2 a + c_{y1}a) $ & I & & & &   \\
online learning variant. \citep{sun2010local} Local-learning updates& \cellcolor{blue!20}$+c_{y2}a $ &&&&&   \\
 between iterations for improved convergence. (\textit{Iterative Relief})& \cellcolor{blue!20} &&&&&   \\

\rowcolor{gray!30} \textbf{TuRF} \citep{moore2007tuning} Address noise and large feature spaces with & \cellcolor{yellow!50}$\mathcal{O}(\mathcal{I} \cdot n^2\cdot a)$ & I & & & &  \\ 
\rowcolor{gray!30}   iterative removal of fixed percent of lowest scoring features. & \cellcolor{blue!20}$\sum_{i=0}^{\mathcal{I}}\big[f(\text{ReliefA})+ $ & & & & & \\
\rowcolor{gray!30}  (\textit{ReliefA}) & \cellcolor{blue!20}$c_{y}a_{i}\log(a_{i})\big]$ & & & & &  \\

\textbf{Evaporative Cooling ReliefF} \citep{mckinney2007evaporative} Address noise and large & \cellcolor{yellow!50}$\mathcal{O}(\mathcal{I} \cdot n^2\cdot a)$ & I & X & & &  \\ 
 feature spaces with iterative `evaporative' removal of $x$ lowest& \cellcolor{blue!20}$\sum_{i=0}^{\mathcal{I}}\big[f(\text{ReliefA})+ $ & & & & &   \\
 quality features via (\textit{ReliefA}) and mutual information (or & \cellcolor{blue!20}$ c_{y1}6an +$ &&&&&   \\
random forest scores) \citep{mckinney2009capturing}. Privacy protection variant \citep{le2017differential}.& \cellcolor{blue!20}$ c_{y2}a_{i}\log(a_{i})\big]$ &&&&&   \\

\rowcolor{gray!30} \textbf{EReliefF} \citep{park2007extended} Seeks to address issues related to incomplete& \cellcolor{yellow!50}$\mathcal{O}(n^2\cdot a)$ & D & X & X & & X \\ 
\rowcolor{gray!30}  and/or multi-class data. (\textit{ReliefF}) & \cellcolor{blue!20} $\chi + c_{3}((p\log p+ q \log q)+ na)$ & & & & & \\

\textbf{VLSReliefF} \citep{eppstein2008very,lee2015very} Efficient interaction detection in large& \cellcolor{yellow!50}$\mathcal{O}(S \cdot n^2\cdot a_{s})$ & E & & & &  \\ 
  feature spaces by scoring random feature subsets. (\textit{ReliefA}) & \cellcolor{blue!20} $S (\chi_{s} + c_{3}2kna_{s} + c_{y1}) +c_{y2}a_{s}$ & & & & &   \\

 \rowcolor{gray!30} \textbf{$i$VLSReliefF} \citep{eppstein2008very} Iterative, TuRF-like extension of & \cellcolor{yellow!50}$\mathcal{O}(\mathcal{I} \cdot S \cdot n^2\cdot a_{s})$ & I & & & &  \\ 
 \rowcolor{gray!30} (\textit{VLSReliefF \& TuRF}). & \cellcolor{blue!20}$\sum_{i=0}^{\mathcal{I}}\big[f(\text{VLSReliefF})+c_{y}a_{i}\log(a_{i})\big] $ & E & & & &   \\

\textbf{ReliefMMS} \citep{chikhi2009reliefmss} Feature weight relative to average feature & \cellcolor{yellow!50}$\mathcal{O}(n^2\cdot a)$ & C & & & &  \\ 
 \emph{diff} between instance pairs. (\textit{ReliefA}) & \cellcolor{blue!20} $\chi + c_{3} 2kna $ & & & & &   \\

\rowcolor{gray!30} \textbf{SURF} \citep{greene2009spatially} Threshold-based nearest neighbors for scoring.& \cellcolor{yellow!50}$\mathcal{O}(n^2\cdot a)$ & C & & & &  \\ 
\rowcolor{gray!30} (\textit{ReliefA \& Iterative Relief})& \cellcolor{blue!20} $\chi + c_{3} 0.5n^2 a $ & & & & &  \\

\textbf{SURF*} \citep{greene2010informative} Introduces `far' scoring to improve detection of  & \cellcolor{yellow!50}$\mathcal{O}(n^2\cdot a)$ & C & & & &  \\ 
 epistatic interactions. (\textit{SURF})& \cellcolor{blue!20} $\chi + c_{3} n^2a $ & & & & &   \\

\rowcolor{gray!30} \textbf{SWRF*} \citep{stokes2012application} Extends (\textit{SURF*}) with sigmoid weighting  & \cellcolor{yellow!50}$\mathcal{O}(n^2\cdot a)$ & C & & & &  \\ 
\rowcolor{gray!30}   taking distance from threshold into account. Introduces & \cellcolor{blue!20}$\chi + c_{3} n^2a + c_{y} 0.25n^2$ & & & & &  \\
\rowcolor{gray!30}   modular framework for Relief development (MoRF).  & \cellcolor{blue!20} & & & & &  \\

 \textbf{LH-RELIEF}  \citep{cai2012feature}  Feature weighting by measuring the margin & \cellcolor{yellow!50}$\mathcal{O}(\mathcal{I} \cdot n^2\cdot a)$ & C & X & X & &  \\ 
 between the sample and its hyperplane. (\textit{I-RELIEF}) & \cellcolor{blue!20} $\chi + c_{3} 2na + c_{y} na$& I & & & &  \\

\rowcolor{gray!30} \textbf{MultiSURF*} \citep{granizo2013multiple} Target instance defined neighbor threshold & \cellcolor{yellow!50} $\mathcal{O}(n^2\cdot a)$ & C & & & &  \\ 
\rowcolor{gray!30} and dead-band no-score zone. (\textit{SURF*}) & \cellcolor{blue!20} $\chi + c_{3} 0.62n^2a + c_{y} n^2$ & & & & &  \\

 \textbf{ReliefSeq}  \citep{mckinney2013reliefseq} Feature-wise adaptive \emph{k}. Choice of $k$ impacts & \cellcolor{yellow!50} $\mathcal{O}(n^2\cdot a)$ & C & X & X & X &  \\ 
differential detection of main effect vs. interaction. (\textit{ReliefA})& \cellcolor{blue!20} $\chi + k_{max}(c_{3} 2k n a) + c_{y} k_{max} \log k_{max} $  & & & & &  \\
 
\rowcolor{gray!30} \textbf{MultiSURF}  \citep{urbanowicz2017rebate} Removed `far' scoring from (\textit{MultiSURF*}) & \cellcolor{yellow!50} $\mathcal{O}(n^2\cdot a)$ & C & X & X & X & X \\ 
\rowcolor{gray!30} to recover main effects. Added strategies to address data  & \cellcolor{blue!20} $\chi + c_{3} 0.31n^2a + c_{y} n^2$ & D & & & &  \\
\rowcolor{gray!30}  types as part of ReBATE. RBAs succeed with heterogeneity.& \cellcolor{blue!20} & & & & &  \\

\hline 
\multicolumn{7}{l}{$\bullet$ $\chi = c_{0}a  +  c_{1} 0.5 n^2 a  + c_{2} n \log n$ \textit{(The universal terms of RBA complete time complexities are given by $\chi$)}}\\
\multicolumn{7}{l}{$\bullet$ \textit{$n$ is the number of training instances, $a$ is the number of features, and $c_{y}$ represents unique constant terms.}} \\
\multicolumn{7}{l}{$\bullet$ $k$ is the user specified number of nearest neighbors; $p$ and $q$ specify the number of hits and misses, respectively.} \\
\multicolumn{7}{l}{$\bullet$ $\mathcal{I}$ is the number of iterations (determined by user parameters) } \\ 
\multicolumn{7}{l}{$\bullet$ $f(\text{ReliefA})$ and $f(\text{VLSReliefF})$ are the complete time complexities of ReliefF and VLSReliefF, respectively) } \\
\multicolumn{7}{l}{$\bullet$ $a_{i}$ is the remaining number of features at iteration $i$} \\ 
\multicolumn{7}{l}{$\bullet$ $a_{i} = a(1-p)^i$ (for TuRF or iVLSReliefF), and $a_{i} < a$ (for Evap. Cool. ReliefF)} \\ 
\multicolumn{7}{l}{$\bullet$ $a_{s}$ is the number of features in each subset, $S$ is the number of subsets, and $\chi_{s}$ is $\chi$ where $a$ is replaced by $a_{s}$} \\ 
\multicolumn{7}{l}{$\bullet$ $k_{max}$ is the maximum $k$ considered by ReliefSeq} \\ 

\end{tabular}
} 
\end{table*}

\subsection{Organizing RBA Research}
Following ReliefF, a number of variations and improvements have been proposed. Table \ref{tab:RBM} chronologically organizes summary information on key RBAs dealing with fundamental feature selection problems. Brief descriptions of the algorithms and their contributions are given along with our designation of the closest parent algorithm in parentheses. Parent algorithms may deviate from what was described in the respective publication(s). This is due to inconsistent nomenclature (e.g. ReliefA implementations being more generically referred to as ReliefF even if they did not include extensions for missing or multi-class data handling) and some previously missed citations of relevant work. In the sections following this review we will adopt the name `ReliefF' for any Relief algorithm that uses $k$ nearest neighbors, and makes the $m=n$ assumption (regardless of any data type handeling implementations), as has become common in the literature \citep{moore2015epistasis, todorov20166}. 

Table \ref{tab:RBM} also provides asymptotic time complexities $\mathcal{O}$ (highlighted in yellow), to easily compare run time order of magnitude. Additionally, based on algorithmic descriptions in the respective publications, we provide approximations of complete algorithm time complexities (highlighted in blue). These equations provide computational insight into algorithmic differences and reveal more subtle run time differences. Variables are defined below the table. Constants in the table are numbered (e.g. $c_{3}$) according to which additive term in the equation that it corresponds to (e.g. $c_{3}2na$). The term with $c_{0}$ initializes the feature weights to $0.0$. The term with $c_{1}$ calculates all unique pairwise distances between instances. Given the assumption of $m=n$, it is computationally more efficient to pre-compute all pairwise distances rather than on a target by target basis as proposed in the original Relief algorithm. The term wit $c_{2}$ corresponds to finding the nearest instances (or separates nearest from furthest in the case of SURF* and MultiSURF*). The term with $c_{3}$ corresponds with updating the feature weights. Algorithms that require additional terms label the corresponding constants generically with $c_{y}$ or a numbered variation. Note that complete time complexity terms that are universal to all RBAs are represented by $\chi$ within the table. 

\citet{todorov20166} suggested that there are two primary directions of RBA development: (1) strategies for selecting and/or weighting neighbors in scoring (i.e. what we call `core algorithm' developments), and (2) strategies for moving beyond a single pass over the data to `iterative' implementations. In Table \ref{tab:RBM}, the column labeled `Focus' identifies the respective research direction(s) of the corresponding algorithm, going beyond the two suggested by Todorov. These include; (1) `C' for \emph{core algorithm}, i.e. variants impacting a single run through the training data such as variations in neighbor selection or scoring, (2) `I' for \emph{iterative approach}, i.e. variants designed to iteratively apply a core Relief algorithm for multiple cycles through the training data, (3) `E' for \emph{efficiency}, i.e. variants seeking to improve computational efficiency, and (4) `D' for \emph{data type handling}, i.e. variants that seek to address the challenges of different data types including continuous feature values, multi-class endpoints, continuous endpoints (i.e. regression), or missing data values. The remaining columns of Table \ref{tab:RBM} indicate (with `X's') whether the corresponding algorithm explicitly considered or implemented algorithm extensions to handle any of the four data types above, beyond discrete features and binary classes.  

We can make some basic observations from this table. First, relatively little attention has been paid to adapting RBAs to regression problems. Second, the majority of proposed variations have focused on data with discrete-valued features and a binary endpoint. Notably many of these works have been application driven, focusing on feature selection in genomics problems with single nucleotide polymorphisms (SNPs) as features that can have one of three discrete values (0, 1, or 2) and a binary endpoint representing sick vs. healthy subjects \citep{moore2007tuning,mckinney2007evaporative,eppstein2008very,greene2009spatially,mckinney2009capturing,greene2010informative,stokes2012application,granizo2013multiple}. RBAs are particularly appealing in this domain since the number of features ($a$) in respective datasets is typically much larger than the number of available training instances ($n$), and RBAs have a linear time complexity with respect to $a$, but a quadratic time complexity with respect to $n$. However, core RBAs aimed at SNP analysis, such as SURF \citep{greene2009spatially}, SURF* \citep{greene2010informative}, SWRF* \citep{stokes2012application}, and MuliSURF* \citep{granizo2013multiple} were not originally extended to handle other basic data types. Table \ref{tab:RBM} concludes with our recently proposed core algorithm named \emph{MultiSURF} \citep{urbanowicz2017rebate}. MultiSURF performed most consistently across a variety of problem types (e.g. 2-way and 3-way interactions as well as heterogeneous associations) in comparison with ReliefF, SURF, SURF*, MultiSURF* and a handful of other non-RBA features selection methods. The work by \citet{urbanowicz2017rebate} also extended MultiSURF along with ReliefF, SURF, SURF*, and MultiSURF* to handle a variety of different data type issues under a unified implementation framework called ReBATE. 

The following subsections go into greater depth describing notable RBAs that fall into our `core', `iterative', `efficiency' or `data type' categories, as well as peripheral RBA research directions not included in this table.

\subsection{Neighbor Selection and Instance Weighting} \label{sec:core}
\begin{figure}[t!]
	\centering
    \includegraphics[width=0.80\textwidth]{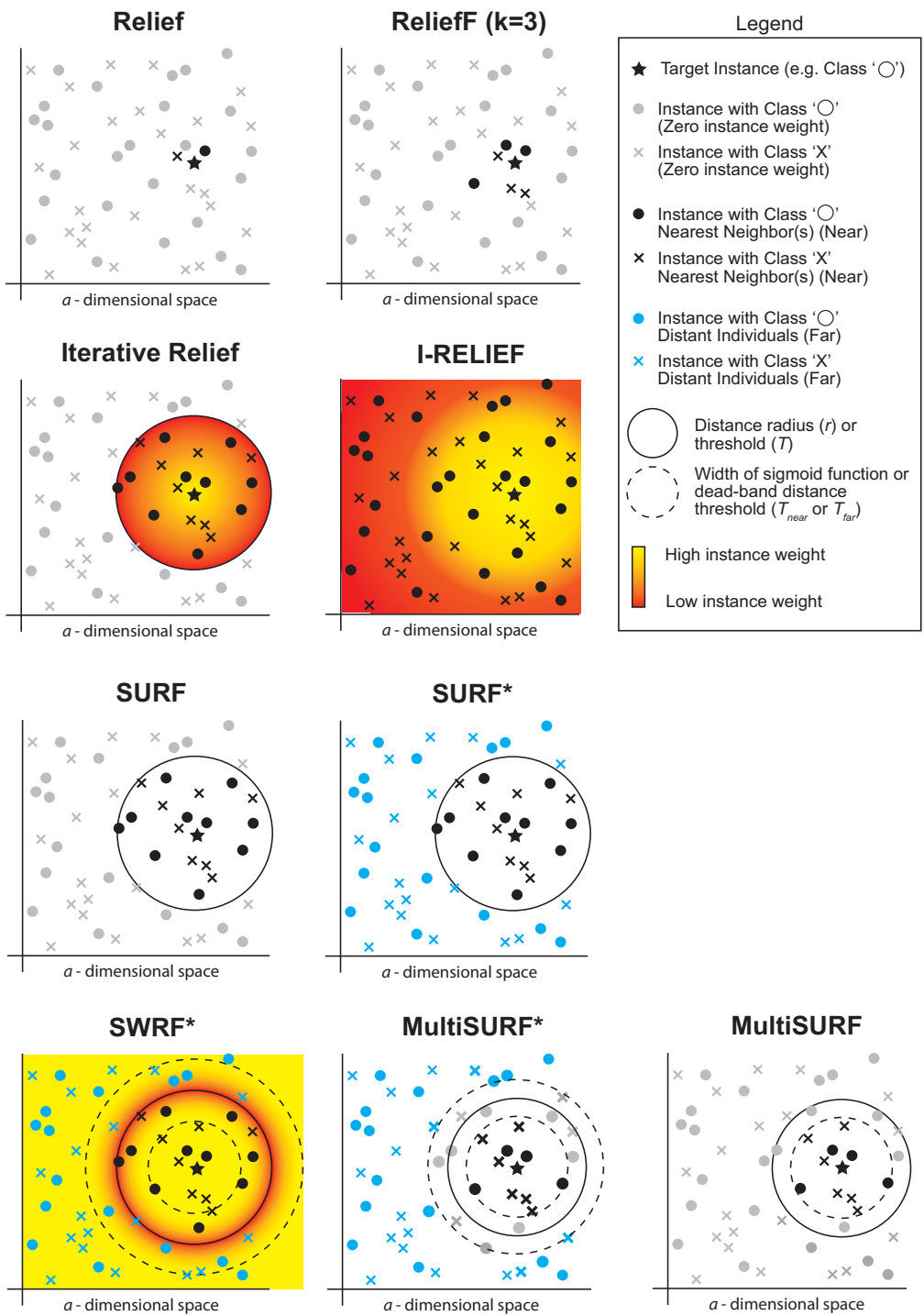}
	\caption{Illustrations of RBA neighbor selection and/or instance weighting schemes. Methods with a red/yellow gradient adopt an instance weighting scheme while other methods identify instances as `near' or `far' which then contribute fully to feature weight updates. These illustrations are conceptual and are not drawn to scale.}

	\label{fig:methods}
\end{figure}

This section references algorithms in Table \ref{tab:RBM} with a core focus (C). How do we select nearest hits and misses? What number of neighboring instances should be used in feature scoring? Is there information to be gained from considering `far' instance pairs? How should the scoring contribution of those neighboring instances be weighted; also referred to as observation weighting by \citet{todorov20166}? These are the primary questions that have been asked in the context of core RBAs. Note that \emph{instance weighting} refers to the weight placed on an instance during the scoring update. By default, most RBAs (including ReliefF) assign neighboring instances a weight of 1, and all others a weight of 0. Figure \ref{fig:methods} illustrates how a variety of RBAs (arranged chronologically) differ with respect to neighbor selection and instance weighting. For every RBA, we assume that each instance in $n$ gets the opportunity to be the target instance during feature scoring. Note that for ReliefF in Figure \ref{fig:methods}, a $k$ of 3 is chosen for illustration simplicity, but a $k$ of 10 is most common. Figure \ref{fig:methods} includes RBAs that adopt a `distance-from-target' instance weighting scheme, i.e. Iterative Relief, I-RELIEF, and SWRF*, where instance weight ranges from 0 to 1. For all other RBAs in the table, instances that are identified as either near or far, have a full weight of 1, while all others have a zero instance weight in feature scoring. Three of the RBAs (i.e. I-RELIEF, SURF*, and SWRF*) are unique in giving all instances, besides the target, some weight each scoring cycle.

 The original Relief algorithm used two nearest neighbors (i.e. one nearest hit and miss), each with an equal instance weighting \citep{kira1992practical} . ReliefA through ReliefF used $k$ nearest neighbors with equal instance weighting \citep{kononenko1994estimating}. Iterative Relief was the first to specify a radius $r$ around the target instance that would define the cutoff for which instances would be considered neighbors \citep{draper2003iterative}. Additionally, while RRelief \citep{kononenko1996relieff} was the first to suggest differentially weighting instances based on their distance from the target instance in regression, Iterative Relief was the first to suggest this for discrete class problems \citep{draper2003iterative}. The effect there was that the closest neighbors had a greater impact on feature weighting than those out towards the edge of the radius. I-RELIEF proposed forgoing the determination of neighbors entirely, instead using an instance weighting function over the entire set of hit and miss instances, again so that the closest neighbors had the greatest impact on feature weighting \citep{sun2006iterative,sun2007iterative,sun2010local}. Similar to Iterative Relief, SURF employed a distance threshold $T$ to define instances as neighbors (where $T$ was equal to the average distance between all instance pairs in the data) \citep{greene2009spatially}. However, in contrast with Iterative Relief, SURF utilizes equal instance weights for all instances defined as neighbors. 
 
 The SURF* expansion introduced the concept of instances that were near vs. far from the target instance \citep{greene2010informative} (see Figure \ref{fig:methods}). Applying the same $T$ from SURF, any instance within the threshold was considered near, and those outside were far. SURF* was similar to I-RELIEF in that all other instances besides the target contributed to scoring. This is reflected in the complete time complexity of the two algorithms where the feature scoring term is $c_{3}n^2a$ for both. However SURF* weights all `near' instances equally, and all `far' instances in a similarly equal, but opposite way. Specifically, for far instances, feature value differences in hits receive a $+1$ while feature value differences in misses receive a $-1$, i.e. the opposite scoring strategy than what is presented in Figure \ref{fig:RS}. Note that in mathematics the '*' indicates opposite, therefore RBAs that utilize 'far' scoring have been given this affix. Some publications have instead have used the affix 'STAR' (e.g. SURFSTAR).
 
 SWRF* integrated concepts from SURF* and I-RELIEF, preserving the definition of near and far established in SURF*, but adopting a sigmoid instance weighting function from I-RELIEF so that the nearest of neighbors have the greatest \emph{standard} scoring weight, while the farthest `far' instances have the greatest \emph{opposite} scoring weight. Instances near $T$ have the smallest scoring weights. The width of the SWRF* sigmoid function is proportional to the standard deviation $\sigma$ of all pairwise instance distances. In contrast with SWRF*, MultiSURF* took an alternate approach to discounting instances near $T$ by introducing a dead-band zone on both the near and far side of $T$ (i.e. $T_{near}$ or $T_{far}$) \citep{granizo2013multiple}. Any instances that fell within this `middle' distance zone were excluded from scoring (i.e. neither near or far). Another major difference is that MultiSURF* defined $T$ as the mean pairwise distance between the target instance and all others, as opposed to the mean of all instance pairs in the data. This adapts the definition of near/far to a given part of the feature space. Similarly, the width of the dead-band zone is the standard deviation $\sigma$ of pairwise distances between the target instance and all others. One final difference between MultiSURF* and SURF* is that the `far' scoring logic was inverted to save computational time. Specifically in SURF*, \emph{differences} in feature values in hits yielded a reduction in feature score, and an increase in misses. Since different feature values are expected to be more frequent in far individuals, in MultiSURF*, \emph{same} feature values in hits yielded an increase in feature score, and a decrease in misses. Also recognizing the importance of neighbor selection, ReliefSeq proposed the concept of an adaptive $k$ for all features \citep{mckinney2013reliefseq}. ReliefSeq effectively examines all possible values of $k$ up to a $k_{max}$ and for each feature, picking the $k$ that yields the largest feature weight in the final scoring. While more computationally intensive, the authors claim that varying k on a feature by feature basis provides greater flexibiliy in detecting either main or interaction effects. Notably, ReliefSeq was applied to the analysis of RNA-Seq expression data. Most recently, MultiSURF was proposed, preserving most aspects of MultiSURF* but eliminating the `far' scoring \citep{urbanowicz2017rebate}. This was due to the fact that while `far' scoring improved the detection of 2-way interactions, it also greatly deteriorated the ability of RBAs to detect simple main effect associations. MultiSURF is claimed to balance performance with respect to its (1) ability to detect main or interaction effects, (2) computational efficiency, (3) ease of use (i.e. no parameters to set), and (4) applicability to a variety of data types.

\subsection{Iterative and Efficiency Approaches} \label{sec:iterative}
This section references algorithms in Table \ref{tab:RBM} with an iterative (I) or efficiency (E) focus. As noted earlier, core RBA performance is understood to degrade as the number of irrelevant features becomes `large' particularly with respect to noisy problems. This has been observed or noted in a number of works \citep{robnik2003theoretical,sun2006iterative,sun2007iterative,moore2007tuning,eppstein2008very,greene2009spatially,todorov20166}. As pointed out by \citet{sun2006iterative}, this is because a core RBA defines nearest neighbors in the original feature space, which are highly unlikely to be the same in weighted space (i.e. the space where we have assigned low weights to features least likely to be relevant). To deal with this issue, iterative and efficiency approaches have been proposed that are wrapped around or integrated into core RBAs. 

\begin{figure}[t!]
	\centerline{\includegraphics[width=0.8\textwidth]{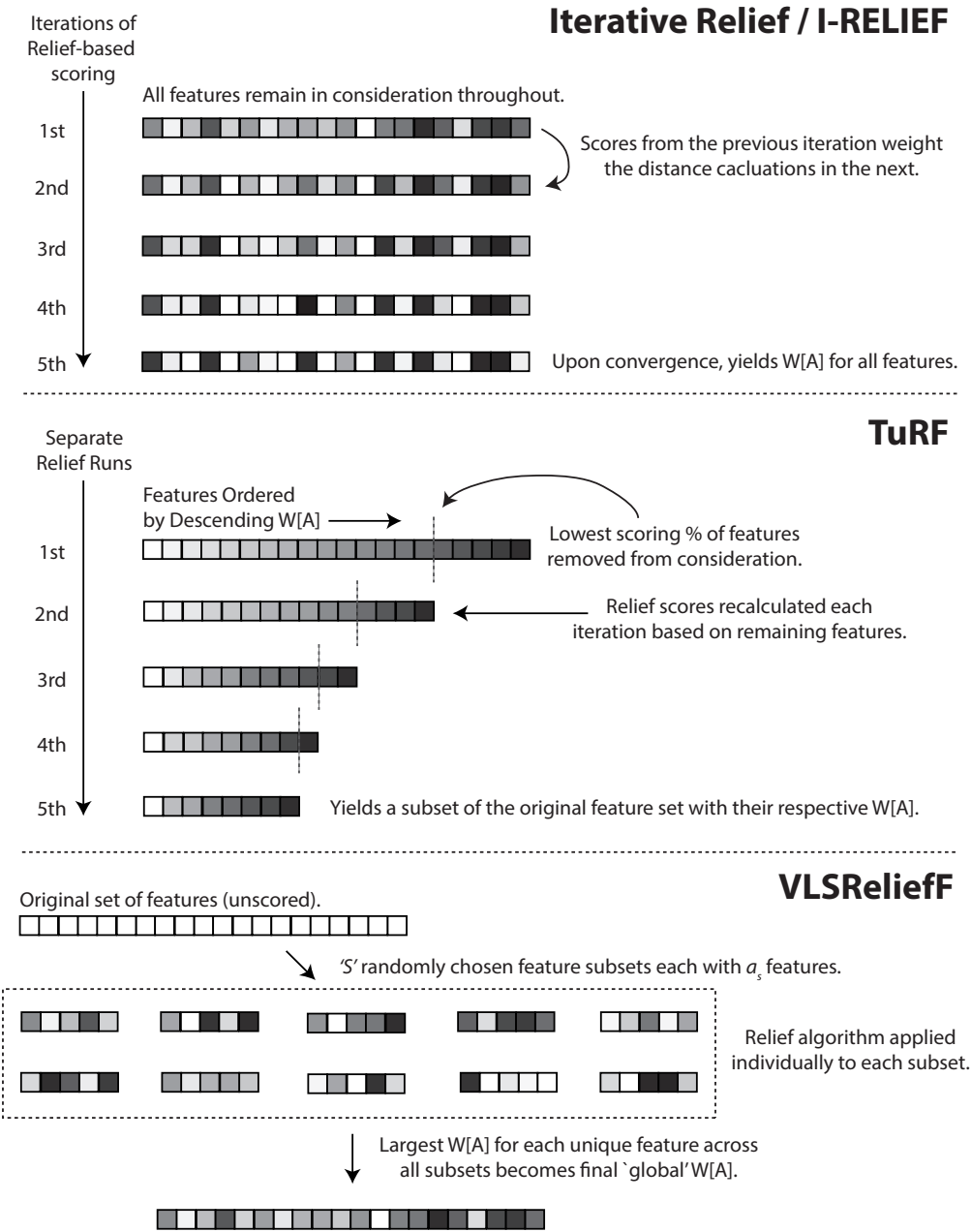}}
	\caption{Illustrations of the basic concepts behind key iterative and efficiency approaches including TuRF, Iterative Relief/I-RELIEF, and VLSReliefF. Features are represented as squares, where darker shading indicates a lower feature weight/score. }
	\label{fig:iterative}
\end{figure}

Iterative Relief introduced the idea of running the core RBA more than once, each time using the feature weights $W$ from the previous iteration to update pairwise distance calculations such that a low scoring feature from the previous iteration has less influence on instance distance in the current iteration \citep{draper2003iterative} (see Figure \ref{fig:iterative}). These `temporary' feature weights were referred to as \emph{parameters} by \citet{todorov20166} and designated by the variable $\phi$. Iteratively updating the distance weights can cause certain samples to enter and leave neighborhoods of other samples. To reduce discontinuities in the feature weight estimates that arise from changing neighborhoods, Iterative Relief also introduced a radius to define neighborhoods rather than a set number of instances, as illustrated in Figure~\ref{fig:methods}. Iterations continued until the weights $W$ converge, or until some maximum number of iterations is reached. It is important to be aware of stop criteria since iterative approaches can become quite computationally expensive. 

Sun and Li independently introduced another iterative Relief method known as I-RELIEF~\citep{sun2006iterative}. I-RELIEF adopts an iterative approach similar to Iterative Relief, but mathematically derived Relief as an online algorithm that solves a convex optimization problem with a margin-based objective function \citep{sun2006iterative,sun2007iterative}. As such, I-RELIEF has been described as an outlier removal scheme since the margin averaging is sensitive to large variations \citep{cai2012feature}. Later, Local Learning I-RELIEF (our name for the unnamed algorithm) applied the concept of local learning to improve iterative convergence by promoting sparse feature weighting \citep{sun2010local}. `Sparse' refers to there being a minimal number of converged feature weights with a value greater than zero. This was achieved by introducing the $\ell_1$ norm penalty (as in \emph{lasso}) into optimization of I-RELIEF. 

TuRF presents a much simpler iterative approach that can easily be wrapped around any other core RBA despite the fact that it was originally designed to be used with ReliefF \citep{moore2007tuning} (see Algorithm \ref{alg:TuRF}). TuRF is essentially a recursive feature elimination approach.  Each iteration, the lowest scoring features are eliminated from further consideration with respect to both distance calculations and feature weight updates (see Figure \ref{fig:iterative}). However selecting the number of iterations ($p$) is not trivial. Evaporative Cooling ReliefF offers another novel approach that employs simulated annealing to iteratively remove lowest relevance features, where relevance is a function of both ReliefF and (myopic) mutual information scores \citep{mckinney2007evaporative}, or instead ReliefF and transformed random forest importance scores \citep{mckinney2009capturing}. Most recently, the evaporative cooling concept was adapted to the challenge of patient privacy preservation, and was extended for continuous feature analysis (i.e. fMRI network data) \citep{le2017differential}. 

\begin{algorithm}[t]
\caption{Pseudo-code of TuRF algorithm}
\label{alg:TuRF}
\begin{algorithmic}
\STATE $\textit{a} \gets$ number of attributes (i.e. features)
\STATE \textbf{Parameter:} $\textit{p} \gets$ number of iterations \\
\STATE
\FOR{i:=1 \TO $p$} 
    \STATE run ReliefF and estimate feature weights ($W$)
    \STATE sort features by weight
    \STATE remove $p/a$ of remaining features with lowest weights
\ENDFOR
\RETURN last ReliefF weight estimates for remaining features
\end{algorithmic}
\end{algorithm}

It was noted by \citet{todorov20166}, that for TuRF, or any of the other iterative approaches that could `remove' features from consideration by giving them a $\phi$ of 0 in the distance calculation, it is still possible to estimate a relevance score \emph{W}[\emph{A}] for it (thus perhaps giving the feature the opportunity to be reintroduced as relevant later). It was warned that this could lead to undesirable oscillatory behavior and poor convergence of scoring. It should also be mentioned that any of the iterative strategies for updating parameters (e.g. I-RELIEF) could also be combined with a specific core RBA besides the one it was originally implemented with (e.g. the iterative component of Iterative Relief could be wrapped around the core SWRF* approach).

Despite the fact that core Relief algorithms are relatively fast, they can still be slow in very large feature spaces (common to bioinformatics), or more importantly, when large training sets are available (because RBAs scale quadratically with the number of instances). One of the more unique RBA proposals focuses on improving algorithm efficiency with regards to both run time and performance. Specifically, VLSReliefF targets the detection of feature interactions in very large feature spaces \citep{eppstein2008very} (see Figure \ref{fig:iterative}). The principle behind VLSReliefF is simply that weights estimated by ReliefF are more accurate when applied to smaller feature sets. Therefore it individually applies ReliefF to some number of randomly selected feature subsets ($S$), each of size $a_{s} < a$ with the expectation that at least one subset in the population will contain all interacting features that are associated with endpoint (and will thus have elevated weights for those features). The partial ReliefF results are integrated by setting the `global' feature weight \emph{W}[\emph{A}] to the maximum `local' weight for a given feature across all $S$ ReliefF runs. With regards to detecting feature interactions, the risk of this approach is that if all relevant interacting features don't appear together in at least one of the $S$ random feature subsets, then the interaction will likely be missed. That is why properly setting the $S$ and $a_{s}$ is critical to maximizing the probability of success. Furthermore, knowing the desired order of interaction to be sought (e.g. 2-way, 3-way) is needed to calculate $S$. The VLSReliefF concept was inspired by work proposing a Random Chemistry ReliefF algorithm, an iterative approach that ran ReliefF on random feature subsets \citep{eppstein2007genomic}. The VLSReliefF concept could be integrated with other core RBAs. An iterative TuRF-like version called $i$VLSReliefF has also been proposed \citep{eppstein2008very}.  

\subsection{Other Relief-based methods}
This section references algorithms in Table \ref{tab:RBM} with a data type focus (D). In the interest of breadth, it also summarizes ancillary Relief expansions not included in Table \ref{tab:RBM}. In a number of studies, emphasis has been placed on the handling of different data types beyond discrete features and binary classes. Beyond RReliefF \citep{kononenko1996relieff,robnik1997adaptation} little attention has been paid to handling continuous endpoints (i.e. regression) other than examples like FARelief \citep{bins2000feature}, or RM-RELIEF \citep{li2014relief}. Alternative methods of handling continuous features beyond those originally introduced in Relief were described by \citet{demvsar2010algorithms} and \citet{blessie2011relief}. An alternative method for handling a multi-class endpoint was described by \citet{ye2008multi}. Little else has been proposed for handling multi-class endpoints or missing data beyond adaptations of those from the original ReliefF algorithm \citep{kononenko1994estimating}.

The Relief concept has also been adapted to a variety of specific data problems.  The most common problem is the removal of redundant features as discussed earlier. Previously, a handful of stand-alone RBAs, or some combination of an existing RBA with a redundancy removal heuristic have been proposed to deal with feature redundancy \citep{bins2001feature,florez2002reviewing,guyon2003multivariate,guyon2005multivariate,yang2006orthogonal,chang2010generalized,mhamdi2013new,zeng2013feature,liu2015feature,challita2015new,agre2016weighted}. Another popular area of investigation is the adaptation of Relief to multi-label learning, i.e. where instances can each have more than one class label assigned to it \citep{kong2012multi,spolaor2012filter,spolaor2013relieff,slavkov2013extending,pupo2013relieff,reyes2015scalable}. Other problems to which RBAs have been adapted include: multiple instance learning, i.e. bags of not clearly labeled instances \citep{zafra2010feature,zafra2012relieff}, dealing with non-monotonic relationships \citep{bins2002evaluating,draper2003iterative}, dealing with survival data (i.e. data exploring the duration of time until one or more events happen) \citep{beretta2011implementing}, dealing with imbalanced data \citep{robnik2003experiments}, clustering \citep{dash2011relief}, and feature extraction \citep{sun2008relief}.  

Other notable Relief methodological variations include approaches for feature set evaluation \citep{arauzo2004feature}, instance selection \citep{dash2007extrarelief}, and ensemble learning \citep{saeys2008robust,zhou2014stable}. Attempts at parallelizing RBAs for run time efficiency have been proposed by \citet{lee2015very} and \citet{eiras2016multithreaded}. Many other works applying RBAs or drawing inspiration from them exist in the literature but are beyond the scope of this methodological review. Earlier reviews in the form of book chapters include (1) Kononenko and Sikonja's focused examination of their own ReliefF and RReliefF contributions, (2) Moore's brief review of ReliefF and select RBAs in the context of epistasis analysis, and (3) Todorov's more recent summarial overview of target RBAs and advancements in the context of detecting gene-environment interactions \citep{kononenko2008non,moore2015epistasis,todorov20166}. 

\subsection{RBA Evaluations} \label{evals}
The datasets chosen to test, evaluate, and compare RBAs in previous studies have often focused on (1) a small sample of simulated or toy benchmark datasets \citep{kira1992practical,chikhi2009reliefmss}, (2) a set of real-world benchmarks (e.g. from the UCI repository) \citep{bins2001feature,florez2002reviewing,qamar2012relief,song2013fast,gore2016feature}, (3) some real data analysis that is new or yet to be established as a benchmark \citep{dessi2013comparative}, or (4) some combination of these three \citep{kononenko1994estimating,robnik2003theoretical,sun2006iterative,sun2007iterative,mckinney2007evaporative,sun2010local,cai2012feature,agre2016weighted,dorani2018feature}.

Some RBAs have been compared across a spectrum of simulated datasets capturing a greater breadth of problem scenarios. This was true for TuRF, SURF, SURF*, SWRF*, and MulitSURF* each developed with the bioinformatic detection of epistastic interactions in mind \citep{moore2007tuning,greene2009spatially,greene2010informative,stokes2012application,granizo2013multiple}. In each of these studies, RBAs were evaluated on datasets with purely epistatic 2-way interactions (i.e. no main effects) with varying numbers of training instances (e.g. 200 to 3200) as well as different heritabilities (e.g. 0.01 to 0.4). Heritability is a genetics term that indicates how much endpoint variation is due to the genetic features.  In the present context heritability can be viewed as the signal magnitude, where a heritability of 1 is a `clean' dataset (i.e. with the correct model, endpoint values will always be correctly predicted based on feature values), and a heritability of 0 would be a completely noisy dataset with no meaningful endpoint associations. All features were simulated as single nucleotide polymorphisms (SNP) that could have have a discrete value of (0, 1, or 2) representing possible genotypes.  In each dataset, two features were predictive (i.e. relevant) of a binary class while the remaining 998 features were randomly generated, based on genetic guidelines of expected genotype frequencies, yielding a total of 1000 features.  Similarly, VLSRelief explored SNP simulations and 2-way epistasis varying heritability similar to the other studies, but fixing datasets to 1600 instances and simulating datasets with either 5000 or 100,000 total features \citep{eppstein2008very}. It should be noted that most of these studies sought to compare core RBAs to respective iterative TuRF expansions, which is why larger feature spaces were simulated.

Another recent investigation compared ReliefF, TuRF, SURF, chi-square, logistic regression, and odds ratio, in their ability to rank features in SNP data simulated to include 15 epistatic feature pairs each contributing additively to class determination \citep{dorani2018feature}. RBAs again performed best both in this simulation, and in identifying interacting SNPs from a real world genome-wide association study (GWAS), confirmed by exhaustive calculation of information gain.

Beyond the simulated genetic analyses described above, there are only a couple examples of comparative evaluations of RBAs over a reasonably diverse set of synthetic datasets including one of Relief \citep{belanche2011review} and another of ReliefF \citep{bolon2013review} in comparison with other feature selection approaches. Notably in both studies, the selected RBA stood out as the more reliable and successful feature selection algorithm, except when dealing with removing feature redundancy. Most recently, a much wider comparison of core RBA algorithms was completed over a broad spectrum of simulated datasets with various properties and underlying patterns; including main effects, interactions, and patterns of genetic heterogeneity \citep{urbanowicz2017rebate}. That study (1) confirmed the utility of RBA methods over chi-square, ANOVA, mutual information, and random forest based approaches for feature selection, (2) illustrated performance differences between a number of core RBAs (i.e. ReliefF, SURF, SURF*, MultiSURF*), and (3) introduced MultiSURF and novel implementations of ReliefF. 

Clearly, the ultimate goal in developing feature selection methods is to apply them to real world problems and ideally facilitating the modeling of previously unknown patterns of association in that data. However, as similarly argued by \citet{robnik2003theoretical}, \citet{belanche2011review}, \citet{bolon2013review}, and \citet{olson2017pmlb}: to properly evaluate and compare methodologies, diverse simulation studies should first be designed and applied.  This is because: (1) Uniquely, a simulation study can be designed by systematically varying key experimental conditions, e.g. varying noise, the number of irrelevant features, or the underlying pattern of association in the data. This allows us to explicitly identify generalizable strengths and weakness of the methods and to draw more useful conclusions; (2) The ground-truth of the dataset is known, e.g. we know which features are relevant vs. irrelevant, we know the pattern of association between relevant features and endpoint, and we know how much signal is in the dataset (i.e. so we know what testing accuracy should be achievable in downstream modeling). This knowledge of ground truth allows us to perform power analyses over simulated dataset replicates to directly evaluate the success rate of our methodologies.  

\subsection{Software Availability}
ReliefF \citep{kononenko1994estimating} and its counterpart for dealing with regression data, i.e. RReliefF \citep{kononenko1996relieff}, are currently the most widely implemented RBAs. They can be found in the following freely available data mining software packages: CORElearn \citep{robnik2012corelearn} (in C++), Weka \citep{hall2009weka} (in Java), Orange \citep{demvsar2013orange}, and R \citep{ihaka1996r} (within the dprep and CORElearn packages). A C++ version of ReliefF is available as part of Evaporative Cooling ReliefF\footnote{\url{https://github.com/insilico/EC/blob/master/src/library/ReliefF.cpp}}.  

Separately, implementations of ReliefF \citep{kononenko1994estimating}, SURF \citep{greene2009spatially}, SURF* \citep{greene2010informative}, and MultiSURF* \citep{granizo2013multiple} as well as the iterative TuRF algorithm \citep{moore2007tuning} were made available in the open source Multifactor Dimensionality Reduction (MDR) \citep{ritchie2001multifactor} software package\footnote{\url{http://sourceforge.net/projects/mdr}}. These Java implementations are computationally efficient, but can only handle `complete data' (i.e. no missing values) with discrete features and a binary endpoint. Python 2.7 versions of these algorithms were later implemented and made available within the open source Extended Supervised Tracking and Classifying System (ExSTraCS)\footnote{\url{https://github.com/ryanurbs/ExSTraCS_2.0}} \citep{urbanowicz2014extended,urbanowicz2015exstracs}. These implementations were less computationally efficient, but extended each algorithm to handle different data types, including continuous features, multi-class endpoints, regression, and missing data. Other C$\#$ implementations of ReliefF, SURF*, and SWRF* were made available as part of the modular framework for Relief development (MoRF)\footnote{\url{https://github.com/mattstokes42/MoRF}} \citep{stokes2012application}. A C$\#$ implementation of ReliefSeq\footnote{\url{http://insilico.utulsa.edu/ReliefSeq}} is also available \citep{mckinney2013reliefseq}. Most recently, ReliefF, SURF, SURF*, MultiSURF*, MultiSURF, and TuRF were all implemented within the Relief-Based Algorithm Training Environment (ReBATE). These ReBATE implementations were coded more efficiently in Python (2 and 3) and similarly extended to handle the aforementioned data types. Stand-alone ReBate software\footnote{\url{https://github.com/EpistasisLab/ReBATE}} and a scikit-learn \citep{scikit-learn} compatible format\footnote{\url{https://github.com/EpistasisLab/scikit-rebate}} were both recently made available.
 
\section{Conclusion} \label{conclusions}
In this work we have placed Relief-based algorithms (RBAs) in the context of other feature selection methods, provided an in-depth introduction to the Relief-algorithm concept, described four general branches of RBA research, and reviewed key methodological differences within these branches.  This work highlights a number of conclusions that can be made about RBAs, including (1) they are generally proficient at detecting not only univariate effects but 2-way interactions as well, (2) they scale linearly well with the number of features, but quadratically with the number of training instances, (3) iterative and efficiency approaches offer a solution to scaling RBAs up very large feature spaces, (4) RBAs are `anytime' algorithms, (5) the choice of instance neighbors is a critical aspect of RBA success, setting these methods apart from other feature selection approaches, (6) the individual feature weights output by an RBA can be used to probabilistically guide downstream machine learning methods (i.e. feature weighting), (7) RBAs have already been flexibly adapted to an array of data types and specific application domains, and (8) implementations of a variety of RBAs are available.

This promising area of feature selection will likely benefit from future research focusing on: (1) the most effective and reliable instance weighting approach (e.g. classic `full' instance weighting, or `distance-from-target-based instance weighting) (2) Optimize the number of neighbors and neighbor selection to optimize RBA performance in a problem-dependent manner (3) improved strategies (e.g. iterative or efficacy) for scaling RBAs to large-scale data (i.e. many features and/or many instances), (4) adapting to other new problem domains (e.g. temporal data), (5) limiting or eliminating user defined RBA run parameters (to make them easier to apply, and require less prior knowledge about the problem domain to set correctly), and (6) new strategies for ensemble feature selection.

RBAs represent a powerful family of feature selection approaches that strike a key balance between ability to detect complex patterns, flexibility to handle different data types, and computational efficiency. While ReliefF has been the staple go-to algorithm of the family for many years, many advancements have since been made. Understanding these advancements is key to selecting the best approach for application as well as in guiding the development of even better feature selection approaches.




\section*{Acknowledgements}
We thank the reviewers for their thoughtful comments. Special thanks to Brian Cole for his constructive feedback. This work was supported by National Institutes of Health grants AI116794, DK112217, ES013508, EY022300, HL134015, LM009012, LM010098, LM011360, TR001263, and the Warren Center for Network and Data Science.



 \bibliographystyle{elsarticle-harv}

\bibliography{refs}

\end{document}